\documentclass[12pt]{article}

\usepackage[numbers,sort&compress]{natbib}
\usepackage{graphicx}
\usepackage{amsmath}
\usepackage{amssymb}
\usepackage{hyperref}
\usepackage[bf]{caption}

\usepackage{sectsty}
\allsectionsfont{\large}

\usepackage{float}

\newcommand{\ba}{\begin{aligned}}
\newcommand{\ea}{\end{aligned}}
\newcommand{\beq}{\begin{equation}}
\newcommand{\eeq}{\end{equation}}
\def\ben{\begin{equation*}}
\def\een{\end{equation*}}
\newcommand{\beqs}{\begin{eqnarray}}
\newcommand{\eeqs}{\end{eqnarray}}

\def\be{\begin{equation}}
\def\ee{\end{equation}}
\def\bea{\begin{eqnarray}}
\def\eea{\end{eqnarray}}
\def\bsp{\be\begin{split}}

\def\a{\alpha}
\def\b{\beta}

\def\e{\epsilon}

\def\s{\sigma}

\makeatletter

\newcommand{\Rmnum}[1]{\expandafter\@slowromancap\romannumeral #1@}
\makeatother

\renewcommand{\title}[1]{\vbox{\center\LARGE{#1}}\vspace{5mm}}
\renewcommand{\author}[1]{\vbox{\center\large{#1}}\vspace{5mm}}

\everymath{\displaystyle}

\begin{document}

\begin{titlepage}
\vspace{10pt} \hfill {ICTP-SAIFR/2013-015} \vspace{20mm}
\begin{center}

{\Large \bf Massless scattering at special kinematics as Jacobi polynomials}

%\vspace{1cm}

\vspace{45pt}

{
\textbf{Chrysostomos Kalousios},$^a$
\footnote[1]{\href{mailto:ckalousi@ift.unesp.br}{\tt{ckalousi@ift.unesp.br}}}
}
\\[15mm]

{\it\ ${}^a\,$ICTP South American Institute for Fundamental Research\\
Instituto de F\'\i sica Te\'orica, UNESP-Universidade Estadual Paulista\\
R. Dr. Bento T. Ferraz 271 - Bl. II, 01140-070, S\~ao Paulo, SP, Brasil}

\vspace{20pt}

\end{center}

\vspace{40pt}

\centerline{{\bf{Abstract}}}
\vspace*{5mm}
\noindent
We study the scattering equations recently proposed by Cachazo, He and Yuan in 
the special kinematics where their solutions can be identified with the zeros of the 
Jacobi polynomials.  This allows for a non-trivial two parameter family of 
kinematics.  We present explicit and compact formulae for the 
$n$-gluon and $n$-graviton partial scattering amplitudes for our special 
kinematics in terms of Jacobi polynomials.  We also provide alternative expressions in terms of gamma functions.  We give 
an interpretation of the common reduced determinant appearing in the amplitudes as the 
product of the squares of the eigenfrequencies of small oscillations of a 
system whose equilibrium is the solutions of the scattering equations.
\vspace{15pt}
\end{titlepage}

\newpage

\section{Introduction}
Recently a new and elegant formula for the complete tree-level S-matrix of pure 
Yang-Mills and gravity in arbitrary dimensions has been given by Cachazo, He and Yuan (CHY) in 
\cite{Cachazo:2013hca}.  It was later extended to a massless colored cubic 
scalar theory \cite{Cachazo:2013iea}.  The formula was proven by Dolan and Goddard for gluon amplitudes in 
\cite{Dolan:2013isa}.  In \cite{Mason:2013sva}, Mason and Skinner constructed a chiral infinite tension limit of the RNS superstring which was shown to compute the CHY formulae.  The chiral infinite tension limit was later generalized to the pure spinor superstring by Berkovits in \cite{Berkovits:2013xba}.  Further progress on the study of the CHY formulae also 
include \cite{Monteiro:2013rya, Adamo:2013tsa, Gomez:2013wza}.

The goal of this work is to study the system of \cite{Cachazo:2013hca} and more 
precisely the scattering equations \cite{Cachazo:2013gna,Cachazo:2012uq}, since 
the latter play an important role in the context of scattering amplitudes.  While the current interest in the scattering equations is certainly the work of CHY, historically they were first written down by Fairlie and Roberts in \cite{Fairlie,Roberts,Fairlie:2008dg} and later in the work of Gross and Mende on the high energy behavior of string theory \cite{Gross:1987ar}.  The have also arisen in the context of twistor string theory in the work of Witten \cite{Witten:2004cp}.

The general case is quite involved and we only focus on a particular kinematic.  
This would allow us to associate the solutions of the scattering equations with 
the zeros of the Jacobi polynomials.  The aforementioned polynomials depend on 
two continuous parameters and so does our kinematics.  After choosing 
polarizations for our system we will be able to obtain compact expressions for 
$n$-gluon and $n$-gravity scattering in arbitrary dimensions.  The final result 
can be written as the product of a piece that  depends on the kinematics and can 
be expressed as Jacobi polynomials and a piece that is related to the 
helicities.

One of the motivations besides 
mathematical curiosity and the desire to obtain explicit results 
for the amplitudes is the fact that systems whose equilibrium is associated 
with the zeros of classical polynomials usually admit a Lax pair 
representation, are completely integrable and in some cases can be explicitly 
solved (employing for example the projection method of Olshanetsky and 
Perelomov \cite{OP}).  Probably the most representative example 
is the Calogero-Moser \cite{Calogero:1975ii, Moser:1975qp} system whose 
equilibrium is the zeros of the Hermite polynomials.

We start in the next section by reviewing the work of  \cite{Cachazo:2013hca} 
and setting up notation.  We then move on to the derivation of our special 
kinematics and also choose appropriate helicities. We continue by evaluating numerically the gluon and gravity amplitudes and give several different but equivalent 
expressions for them.  We then give an interpretation of the common reduced 
determinant appearing in both Yang-Mills and gravity amplitudes and finally we 
present our conclusions.  In an appendix we summarize properties of the roots 
of the Jacobi polynomials.

\section{Prolegomena}
In the work of \cite{Cachazo:2013gna} it was pointed out the existence of polynomial equations that connect the space of kinematic invariants of $n$ massless particles with momentum $k_a$ in arbitrary spacetime dimensions with the positions $\sigma_c$ of $n$ points on a Riemann sphere through  
\be \label{scattering_equations}
\sum_{b=1,\,b\neq a}^n \frac{k_a \cdot k_b}{\s_a-\s_b} = 0.
\ee
Due to some remarkable properties that these equations possess they were proposed to play an important role in the scattering of massless particles and were called the \textit{scattering equations}.  These equations are invariant under $SL(2,\mathbb{C})$ transformations, which allows us to fix three of the $\sigma$s to arbitrary values.  

The connection with the tree level $S$-matrix of massless particles was made clear in the subsequent work of CHY \cite{Cachazo:2013hca}, where it was proposed that the tree level $n$-gluon partial amplitude of Yang-Mills is given by
\be\label{gluon_amplitude} 
A_n = \sum_{\{\s\}~\in~\mathrm{solutions}} \frac{1}{\s_{12}\cdots\s_{n1}}
\frac{\mathrm{Pf}'\Psi(k,\epsilon,\s)}{\mathrm{det}'\Phi}
\ee
and that of gravity by
\be\label{gravity_amplitude} 
M_n = \sum_{\{\s\}~\in~\mathrm{solutions}}
\frac{\mathrm{det}'\Psi(k,\epsilon,\s)}{\mathrm{det}'\Phi},
\ee
where the sum runs over all the $(n-3)!$ solutions of 
\eqref{scattering_equations} and $\s_{ab} = \s_a - \s_b$.  

In order to explain the above notation we start by defining the 
$2n\times 2n$ antisymmetric matrix $\Psi$ as
\be\label{Psi}
\Psi = 
\begin{pmatrix}
A & -C^{\mathrm{T}} \\
C & B
\end{pmatrix}
\ee
with the $n \times n$ matrices $A,~B,~C$ defined as
\be
A_{ab} = 
  \begin{cases}
  \frac{k_a \cdot k_b}{\s_a-\s_b} & \quad a\neq b, \vspace{0.3cm}\\
  0 &\quad a=b,
 \end{cases}
\qquad\qquad
B_{ab} = 
 \begin{cases}
  \frac{\epsilon_a \cdot \epsilon_b}{\s_a-\s_b} & \quad a\neq b, \vspace{0.3cm} \\
  0 &\quad a=b,
 \end{cases}
\ee
\be
C_{ab} = 
 \begin{cases}
  \frac{\epsilon_a \cdot k_b}{\s_a-\s_b} & \quad a\neq b, \vspace{0.3cm} \\ 
  -\sum_{c\neq a} \frac{\epsilon_a \cdot k_c}{\s_a-\s_c} &\quad a=b,
 \end{cases}
\ee
where $\e_a$ are the helicities of the external particles. We further define the $n\times n$ matrix $\Phi$ through
\be \label{Phi}
\Phi_{ab} =
 \begin{cases}
  \frac{k_a \cdot k_b}{(\s_a-\s_b)^2} & \quad a\neq b, \vspace{0.3cm}\\
  -\sum_{c\neq a} \frac{k_a \cdot k_c}{(\s_a-\s_c)^2} &\quad a=b.
 \end{cases}
\ee
Then, the reduce Pfaffian appearing in the numerator of \eqref{gluon_amplitude} is defined to be $\mathrm{Pf}'\Psi = 2\frac{(-1)^{i+j}}{\s_i-\s_j} \mathrm{Pf}(\Psi_{ij}^{ij})$, where the matrix $\Psi_{ij}^{ij}$ is derived from the matrix $\Psi$ by removing the $i$th and $j$th rows and the $i$th and $j$th columns with $1\leq i<j\leq n$.  The numerator of \eqref{gravity_amplitude} is defined $\mathrm{det}'\Psi = 4 \,\mathrm{det}\Psi_{ij}^{ij} / \s_{ij}^2$, whereas the common denominator of \eqref{gluon_amplitude} and \eqref{gravity_amplitude} is given by $\mathrm{det}'\Phi = \frac{|\Phi|^{ijk}_{pqr}}{(\s_{ij}\s_{jk}\s_{ki})(\s_{pq}\s_{qr}\s_{rp})}$, where the minor $|\Phi|^{ijk}_{pqr}$ is the determinant of the matrix $\Phi$ after removing rows $\{i,j,k\}$ and columns $\{p,q,r\}$.

In the following sections we occasionally use the 
short notation  $k_{ab} = k_a \cdot k_b$ and $\e_{ab} = 
\e_a \cdot \e_b$.

\section{Derivation of the special kinematics}
It turns out that we can uniquely choose our kinematics in such a way that the 
solutions to \eqref{scattering_equations} are identified with the zeros of the 
Jacobi polynomials.  We fix the $SL(2,\mathbb{C})$ invariance by choosing $\s_1 
= -1,~\s_2=1,~\s_3=\infty$.  Then for $a\geq4$, equation 
\eqref{scattering_equations} gives
\be\ba\label{jacobi_scattering}
\sum_{b=4,\,b\neq a}^n \frac{k_{ab}}{\s_a-\s_b} = \frac{(k_{a2}-k_{a1})+(k_{a2} + k_{a1}) \s_a}{1-\s_a^2}, \qquad 
a \geq 4.
\ea\ee
We now choose the special kinematics 
$k_{a1}=(1+\b)/2,~k_{a2}=(1+\a)/2,~k_{ab}=1$ for $a,b\geq 4$, so that the 
$(n-3)$ variables $\s_a$ in \eqref{jacobi_scattering} can be interpreted 
according to \eqref{jacobi_roots} as the $(n-3)$ roots of the Jacobi polynomial 
$P_{n-3}^{(\a,\b)}$.  It is known that only $(n-3)$ of the equations in 
\eqref{scattering_equations} are independent, therefore the rest of the 
kinematics can be derived from the conservation of momentum.  As a consistency 
check we can consider the scattering equations \eqref{scattering_equations} for 
$a=1$ and $a=2$.  Using \eqref{scattering_equations} and \eqref{jacobi_sum1} we 
get $k_{12}=(1+\b) P_{n-3}^{(\a,\b)}\,'(-1)/P_{n-3}^{(\a,\b)}(-1)$ and 
$k_{12}=-(1+\a) P_{n-3}^{(\a,\b)}\,'(1)/P_{n-3}^{(\a,\b)}(1)$ respectively.  The 
two values match and equal to $k_{12}=(3-n)(\a+\b+n-2)/2$ in accordance with the 
conservation of momentum.  We summarize our results in the following table
\begin{table}[H]
\centering
\begin{tabular}{l r}\hline\hline
 $k_{12} = (3-n)(\a+\b+n-2)/2$  \\
 $k_{13} = (n-3)(n-3+\a)/2$   \\
 $k_{23} = (n-3)(n-3+\b)/2$  \\
 $k_{1a} = (1+\b)/2$ & $a\geq 4$\\
 $k_{2a} = (1+\a)/2$ & $a\geq 4$\\
 $k_{3a} = (6-2n-\a-\b)/2$ & $a\geq 4$\\
 $k_{ab} = 1$ & $a,b\geq 4, a\neq b $\\
 \hline
\end{tabular}
\caption{Our two parameter special kinematics.}
\label{special_kinematics}
\end{table}

We now want to choose polarization vectors compatible with $\e_a \cdot k_a = 0$ 
for every $a$ as well as  conservation of momentum.  We choose according to the 
criterion that in the limit $\s_3 \to \infty$ the quantity 
$\s_{23}^2\s_{31}^2\mathrm{det}(\Psi_{12}^{12})$ is invariant under $\s_i$ 
permutations with $i\geq 4$.  Since the general case seems to be quite involved 
we also choose to further simplify our problem by making the following choice
\begin{table}[H]
\centering
\begin{minipage}[b]{55mm}
\begin{tabular}{l r}\hline\hline
 $\e_a\cdot k_b = 0$ & $a,b\geq 1$ \\
 \hline
\end{tabular}
\vspace{0pt}
\end{minipage}
\begin{minipage}[b]{55mm}
\begin{tabular}{l r}\hline\hline
 $\e_{12},\e_{13},\e_{23}$ & arbitrary \\
 $\e_{1a}=c_1$ & $a \geq 4$ \\
 $\e_{2a}=c_2$ & $a \geq 4$ \\
 $\e_{3a}=c_3$ & $a \geq 4$ \\
 $\e_{ab}=c_4$ & $a,b \geq 4, a\neq b$ \\
 \hline
\end{tabular}
\vspace{0pt}
\end{minipage}
\caption{Helicity choice.}
\label{helicities}
\end{table}
\noindent
where $c_i$ are arbitrary constants.  We assume spacetime dimension large 
enough compared to the number of particles.

\section{Evaluation of the amplitudes}
In order to evaluate the reduced determinant we choose to remove the first three columns and the first three rows of \eqref{Phi}.  Then the $(n-3) \times (n-3)$ reduced matrix $\Phi$ becomes
\be \label{Phi_reduced}
(\Phi_{ab})^{123}_{123} =
 \begin{cases}
  \frac{1}{(\s_{a}-\s_{b})^2} & \quad a\neq b, \vspace{0.3cm}\\
  -\frac{1+\b}{2(\s_{a}+1)^2} - \frac{1+\a}{2(\s_{a}-1)^2}-\sum_{c=4,\,c\neq a}^n \frac{1}{(\s_{a}-\s_{c})^2} &\quad a=b=4,\ldots, n,
 \end{cases}
\ee
One observes that the determinant of \eqref{Phi_reduced} is independent of 
swapping any $\s_i$ with any $\s_j$.  To see that we have to swap row $(i-3)$ 
with row $(j-3)$ and column $(i-3)$ with column $(j-3)$.  The determinant of 
\eqref{Phi_reduced} is evaluated to 
\be
|\Phi|^{123}_{123}=\frac{\left({P^{(\a,\b)}_{n-3}}^{(n-3)}(x)\right)^3}{(n-3)! P^{(\a,\b)}_{n-3}(1) P^{(\a,\b)}_{n-3}(-1)}\,,
\ee
where the quantity in the numerator means the $(n-3)$th derivative.  We do not have an analytic proof of the above result although a proof based on recurrence relations of the  Jacobi polynomials might be within reach.  We have checked our result numerically for up to $n=20$.

The following sum over all $(n-3)!$ permutations of $\{1,2,\ldots,n-3\}$ is
\be
\sum_{\mathrm{perms}} \frac{1}{\s_{45}\s_{56}\cdots(\s_n+1) }
=\frac{1}{(\s_4+1)(\s_5+1)\cdots(\s_n+1)}
=\frac{(-1)^{n+1}}{(n-3)!} \frac{{P^{(\a,\b)}_{n-3}}^{(n-3)}(x)}{P^{(\a,\b)}_{n-3}(-1)}.
\ee

In order to evaluate $\mathrm{det}'\Psi$ we choose to eliminate the first and second row and column of \eqref{Psi}.  We have found numerically (up to $n$=20) that for odd $n$ the determinant vanishes whereas for even $n$ it is
\be
\s_{23}^2\s_{31}^2\mathrm{det}(\Psi_{12}^{12})
=\left( 
 \frac{2(n-3)!!}{((n-4)!!)^2}
 \frac{{P^{(\a,\b)}_{n-3}}^{(n-3)}(x)    {P^{(\frac{\a-1}{2},\frac{\b-1}{2})}_{\frac{n}{2}-1}}^{(\frac{n}{2}-1)}(x)}
  {P^{(\frac{\a}{2},\frac{\b}{2})}_{\frac{n}{2}-2}(1) P^{(\frac{\a}{2},\frac{\b}{2})}_{\frac{n}{2}-2}(-1)} H_n
\right)^2\,,
\ee
with the helicity dependent part $H_n$ given by
\be
H_n = \frac{c_4^{n/2-3}}{2(n-3)+\a+\b}
\left( \frac{2(n-3)(n-4)c_1c_2c_3}{(1+\a)(1+\b)}-c_3c_4\e_{12}
\right) 
+ c_4^{n/2-2} \left( \frac{c_2 \e_{13}}{1+\a}
                      +\frac{c_1 \e_{23}}{1+\b}
               \right).
\ee
As a consistency check one can verify the vanishing of $H_n$ under the replacement $\e_a \to k_a$ as required by gauge invariance.

Putting all pieces together we arrive at the final expressions for the amplitudes.  For odd $n$ the amplitudes vanish, whereas for even $n$ we find
\be\ba
A_n &= \frac{ (-1)^{n/2}2^{5-n}(n-2)(n-3)!!    P^{(\frac{\a-1}{2},\frac{\b-1}{2})}_{\frac{n}{2}-1}(1)  }
           {  P^{(\frac{\a}{2},\frac{\b}{2})}_{\frac{n}{2}-2}(-1)  {P^{(\frac{2\a+1}{4},\frac{2\b+1}{4})}_{\frac{n}{2}-2}}^{(\frac{n}{2}-2)}(x)  } H_n,\\
M_n &= \frac{ 2^{7-n}(n-2)^2((n-3)!!)^2   P^{(\frac{\a-1}{2},\frac{\b-1}{2})}_{\frac{n}{2}-1}(1)   
             P^{(\frac{\a-1}{2},\frac{\b-1}{2})}_{\frac{n}{2}-1}(-1) 
             {P^{(\frac{\a-1}{2},\frac{\b-1}{2})}_{\frac{n}{2}-1}}^{(\frac{n}{2}-1)}(x)
           }
           { P^{(\frac{\a}{2},\frac{\b}{2})}_{\frac{n}{2}-2}(1)  
             P^{(\frac{\a}{2},\frac{\b}{2})}_{\frac{n}{2}-2}(-1)
             {P^{(\frac{2\a+1}{4},\frac{2\b+1}{4})}_{\frac{n}{2}-2}}^{(\frac{n}{2}-2)}(x)
           }H_n^2.
\ea\ee
We also provide two alternative expressions
\be\ba
A_n &= \frac{2(n-3)!!\prod_{j=1}^{n/2-1}(2j-1+\a)}{\prod_{j=1}^{n/2-2}(2j+\b)(2j+n-3+\a+\b)}H_n ,\\
M_n &= -\frac{2^{5-n}((n-3)!!)^2\prod_{j=1}^{n/2-1}(2j-1+\a)(2j-1+\b)(2j+n-4+\a+\b)}
            {\prod_{j=1}^{n/2-2}(2j+\a)(2j+\b)(2j+n-3+\a+\b)} H_n^2
\ea\ee
and
\be\ba
A_n &= 2^{4-n/2}(n-3)!!\frac{\Gamma\left(\frac{n-1+\a}{2}\right)
            \Gamma\left(1+\frac{\b}{2}\right)
            \Gamma\left(\frac{n-1+\a+\b}{2}\right)
           }
           {\Gamma\left(\frac{1+\a}{2}\right)
           \Gamma\left(\frac{n-2+\b}{2}\right)
           \Gamma\left(\frac{2n-5+\a+\b}{2}\right)
           }H_n,\\
M_n &=-2^{8-n}((n-3)!!)^2 \times\\
    & \frac{\Gamma\left(1+\frac{\a}{2}\right)
            \Gamma\left(\frac{n-1+\a}{2}\right)
            \Gamma\left(1+\frac{\b}{2}\right)
            \Gamma\left(\frac{n-1+\b}{2}\right)
            \Gamma\left(\frac{n-1+\a+\b}{2}\right)
            \Gamma\left(\frac{2n-4+\a+\b}{2}\right)
           }
           {\Gamma\left(\frac{1+\a}{2}\right)
           \Gamma\left(\frac{n-2+\a}{2}\right)
           \Gamma\left(\frac{1+\b}{2}\right)
           \Gamma\left(\frac{n-2+\b}{2}\right)
           \Gamma\left(\frac{n-2+\a+\b}{2}\right)
           \Gamma\left(\frac{2n-5+\a+\b}{2}\right)
           }H_n^2.
\ea\ee

\section{An interpretation of the reduced determinant}
For our special kinematics the scattering equations \eqref{jacobi_scattering} become
\be\label{scattering_equations_special_kinematics}
\sum_{j = 4}^n \frac{1}{\s_i-\s_j} +\frac{1+\b}{2(\s_i+1)} + \frac{1+\a}{2(\s_i-1)} = 0.
\ee
We now consider the $(n-3)$-body dynamical system described by the Hamiltonian
\be\label{system}
H = \frac 1 2 \sum_{i=1}^{n-3} p_i^2 + U
\ee
with potential energy
\be\label{potential} 
U = \sum_{i<j}^{n-3} \ln |x_i-x_j| + \frac{1+\b}{2}\sum_{i=1}^{n-3} \ln |x_i+1| +\frac{1+\a}{2}\sum_{i=1}^{n-3} \ln |x_i-1|.
\ee
One may think of \eqref{system}-\eqref{potential} as a system of $n$ particles, 
one of them at the fixed position $x=-1$ with `charge' 
\footnote{From now on we drop the quotation marks around `charge'.} $(1+\b)/2$, 
the second at $x=1$ with charge $(1+\a)/2$, the third at infinity 
 \footnote{One may be bother by the fact that the interaction of the 
particle at infinity gives rise to a term proportionally to $\ln(\infty)$ in the 
potential evergy.  This is not a problem though, since the sum of all 
interactions of the particle at infinity with the rest $(n-1)$ particles is 
exactly zero, due to the conservation of momentum.} and the remaining $(n-3)$ 
which have unit charge are bound to move in the interval $(-1,\,1)$ 
\footnote{We can specialize to the case $\a>-1$ and $\b>-1$ where all the zeros 
of the Jacobi polynomials are simple and lie in the interval $(-1,1)$.}.  The 
particles of our system preserve their ordering.

Two observations follow.  The equations of motion for the system 
\eqref{system}-\eqref{potential} are
\be
\ddot{x}_i = -\sum_{j = 1}^{n-3} \frac{1}{x_i-x_j} - \frac{1+\b}{2(x_i+1)} - \frac{1+\a}{2(x_i-1)}\,,
\ee
therefore \eqref{system}-\eqref{potential} has an equilibrium at the solutions 
of the scattering equations \eqref{scattering_equations_special_kinematics}.  
Moreover the squares of the eigenfrequencies of small oscillations of 
\eqref{system}-\eqref{potential} around the equilibrium position are given by 
the eigenvalues of the matrix
\be 
\partial_i\partial_j U = \left( -\frac{1+\b}{2(x_i+1)^2} - \frac{1+\a}{2(x_i-1)^2}-\sum_{l=1,\,l\neq j}^{n-3} \frac{1}{(x_i-x_l)^2} \right) \delta_{ij} + \frac{1}{(x_i-x_j)^2}(1-\delta_{ij})\,,
\ee
which is precisely \eqref{Phi_reduced}.  The generalization to general kinematics is obvious.

In \cite{Corrigan:2002th} it was found that the zeros of Jacobi polynomials are 
related to the equilibrium of the $BC_{n-1}$ Sutherland model and a Lax 
representation was given for the special case $\a=\b=1$.  We do not study 
further the system \eqref{system}-\eqref{potential} here.

\section{Concluding remarks}
In this work we have presented the $n$-gluon and $n$-graviton partial amplitudes for a non-trivial two-parameter family of 
kinematics.  After we have chosen convenient helicities for our systems we were 
able to write the gluon and gravity partial scattering amplitudes in an 
explicit and compact form.  The key idea behind this calculation was the 
observation that the scattering equations \eqref{scattering_equations} can be a 
special case of the equations that the zeros of the Jacobi polynomials satisfy.

Although the solutions of the scattering equation associated with the roots of 
Jacobi polynomials are in general complicated, the final result for the 
amplitude is surprisingly simple.  This simplicity is due to several properties 
of the Jacobi polynomials that are expressed as cancellations in the evaluation 
of the amplitudes.  It is possible that other polynomials could also lead to 
other kinematics and simple final expressions.

In \cite{Cachazo:2013iaa} it was shown that for four dimensional kinematics the 
scattering equations possess $(n-3)!$ solutions.  This result was later 
extended to arbitrary dimensions in \cite{Cachazo:2013gna}.  In our case this 
number follows naturally as all possible ways to permute the $(n-3)$ 
distinct solutions of the $(n-3)$th order Jacobi polynomial.  Another 
observation is that the final expressions are simpler than the individual 
pieces, since cancellations occurred when we combined all pieces together.

We have also associated the common reduced determinant appearing in 
\eqref{gluon_amplitude} and \eqref{gravity_amplitude} (and also in the 
colored ordered scalar theory in \cite{Cachazo:2013iea}) with a 
system whose equilibrium is the solutions of the scattering equations. This 
observations holds also for general kinematics.  The simple result of the 
reduced determinant in our kinematics asks for a simple derivation which we do 
not have at the moment.

Perhaps one of the most interesting observation of this work is the existence 
of a Lax pair for the system \eqref{system}-\eqref{potential}, even if this is 
only known for a special case.  This hints to a possible integrable 
structure behind the amplitudes at least at tree level and the possibility to 
study the amplitudes as an $N$-body system.  It would also be interesting to 
find a $Y$-system description for our system possibly along the lines of 
\cite{Alday:2009dv, Alday:2010vh}.

\vspace{3mm}

\noindent
{\bf Acknowledgments}

\vspace{3mm}

\noindent
We thank Nathan Berkovits, Wei He, Francisco Rojas, Stephan Stieberger, Jon Toledo and Pedro Vieira for useful comments and discussions.  We also thank the organizers of \emph{``Program on Amplitudes and Correlation Functions''} at ICTP in S\~{a}o Paulo for creating a stimulating environment during the course of this work.

\setcounter{equation}{0}
\def\theequation{A.\arabic{equation}}

\appendix
\section{Properties of the roots of the Jacobi polynomials}
In this appendix we review some useful properties of the roots of the Jacobi 
polynomials $P_n^{(\a,\b)}(x)$.  The reader may also consult the classical 
reference \cite{Szego:1939} as well as \cite{Ahmed:1978uw}.

The Jacobi polynomials obey the differential equation
\be
(1-x^2)y''(x)+(\b-\a-(\a+\b+2)x) y'(x)+n(n+\a+\b+1) y(x)=0.
\ee

For $\a>-1$ and $\b>-1$ the $n$th order polynomial has $n$ distinct roots that lie in the interval $(-1,\,1)$.

Next we prove some identities associated to the roots of $P_n^{(\a,\b)}(x)$.  We 
start by expressing the Jacobi polynomials as $k\prod_{j=1}^{n} (x-x_j)$.  
After taking the logarithm and then the derivative we arrive at
\be \label{jacobi_sum1}
\sum_{j=1}^n \frac{1}{x-x_j}= \frac{P_n^{(\a,\b)}\,'(x)}{P_n^{(\a,\b)}(x)}
\ee
or
\be \label{jacobi_sum2}
\sum_{j\neq i}^n \frac{1}{x-x_j}= \frac{(x-x_i)P_n^{(\a,\b)}\,'(x)-P_n^{(\a,\b)}(x)}{(x-x_i)P_n^{(\a,\b)}(x)}.
\ee
Taking the limit where $x$ approaches the root $x_i$ and applying de l'H\^{o}pital's rule twice we arrive at
\be\label{jacobi_roots}
\sum_{j\neq i}^n \frac{1}{x_i-x_j}=\frac{\a-\b+(\a+\b+2)x_i}{2(1-x_i^2)}. 
\ee
Similarly, by differentiating \eqref{jacobi_sum2}, taking the limit where x 
approaches the root $x_i$ and applying de l'H\^{o}pital's rule four times we 
arrive at
\be\ba
12 & (1-x_i^2)^2 \sum_{j\neq i}^n \frac{1}{(x_i-x_j)^2}=4(n-1)(\a+\b+n+2)-(\a-\b)^2 \\
& -2(\a-\b)(\a+\b+6)x_i - (4n(\a+\b+n+1)+(\a+\b+2)(\a+\b+6))x_i^2.
\ea\ee

\bibliographystyle{utphys}
%\addtolength{\parskip}{-1ex}
\bibliography{mybib}
\end{document}